\documentclass{svmult}
\usepackage{mathptmx}       
\usepackage{helvet}         
\usepackage{courier}        
\usepackage{type1cm}        
\usepackage{makeidx}         
\usepackage{graphicx}        
\usepackage{multicol}        
\usepackage[bottom]{footmisc}

\makeindex             

\begin{document}

\title*{The Chemical and Dynamical Evolution of Isolated Dwarf Galaxies}
\titlerunning{Chemodynamical Simulations of Dwarf Galaxies}
\author{Kate Pilkington, Brad K. Gibson, Francesco Calura, \break 
Greg S. Stinson, Chris B. Brook and Alyson Brooks}
\authorrunning{Pilkington et~al.} 
\institute{Kate Pilkington \at Jeremiah Horrocks Inst, Univ of Central 
Lancashire, Preston, UK \email{kpilkington@uclan.ac.uk}}

\maketitle

\abstract{Using a suite of simulations \cite{Gov10} which
sucessfully produce bulgeless (dwarf) disk 
galaxies, we provide an analysis of their associated cold 
interstellar media (ISM) and stellar chemical abundance patterns.  
A preliminary comparison with observations is undertaken, in 
order to assess whether the properties of the cold gas and chemistry 
of the stellar components are recovered successfully. To this end, we 
have extracted the radial and vertical gas density profiles, neutral 
hydrogen velocity dispersion, and the power spectrum of structure within 
the ISM. We complement this analysis of the cold gas with a brief 
examination of the simulations' metallicity distribution functions and 
the distribution of $\alpha$-elements-to-iron.}

\section{Introduction}
\label{intro}

\vspace{-5mm}
Historically, simulating the formation and evolution of a bulgeless, 
rotating, stellar disk, within the context of the classical picture of 
hierarchical assembly, has proven problematic \cite{Gov07}.  Overcoming 
this has only been achieved recently \cite{Gov10}, with 
Governato et~al. presenting simulated 
dwarfs with bulge-to-disk ratios $<$0.04, with very shallow associated 
central dark matter profiles.  This was achieved by resolving the 
inhomogeneous ISM, thereby allowing for the imposition of a realistic 
star formation density threshold which resulted in strong 
supernovoae-driven outflows preferentially removing low angular 
momentum gas.

In what follows, we concentrate our analysis on three simulated dwarf 
galaxies, the basic properties for two of 
which (DG1, DG1LT) have been introduced previously \cite{Gov10}. In 
\S\ref{sec:2}, we quantify the spatio-kinetic characteristics of these 
bulgeless dwarfs, and contrast them with recent observational work 
\cite{Obr10,Stan99,Tam09}.  A preliminary analysis of the chemical 
properties of the associated stellar components is then presented in 
\S\ref{sec:3} and compared with relevant empirical work 
in the field \cite{Scho09,Tol09}. As the simulations are not modelled on any
one particular galaxy, the analysis should be seen as comparative only
in a qualitative sense.

\subsection{Simulations}
\label{simulations}

\vspace{-5mm}
The three simulations analysed were run using the N-body$+$SPH 
code \textsc{gasoline} \cite{Wad04}, employing the same initial 
conditions, and differing only in the adopted resolution, star formation 
density threshold, feedback efficiency, and star formation efficiency. 
The base simulation (DG1) \cite{Gov10} uses a star formation density 
threshold a thousand times higher (100~cm$^{-3}$) than that adopted in 
previous cosmological simulations ($\sim$0.1~cm$^{-3}$); a lower density 
threshold version (DG1LT), despite its possession of classical 
problematic traits (i.e., centrally-concentrated light (B/D$\approx$0.3) 
and dark matter density distributions), is included here for 
comparison.  This implementation of a more realistic density threshold 
for star formation was made possible by the associated high force 
resolution (86pc) which allowed us to resolve individual star forming 
regions of mass $\sim$10$^5$~M$_\odot$.  We include in our analysis a 
new simulation (nDG1), similar to DG1, but with two important
modifications - the inclusion of high-temperature (T$>$10$^4$K) 
metal-line cooling \cite{SWS10}, and enhanced supernova feedback (100\% 
of the energy being coupled to the surrounding ISM in the form of 
thermal energy, rather than the 40\% used in the base simulation). 
The feedback method is described fully in 
\cite{Stinson06}.

\section{HI Analysis of the Simulated Dwarfs} 
\label{sec:2} 

\vspace{-5mm}
Throughout our analysis, we cross-check results which are based on 
cold gas particles (T$<$15000K) \cite{Stinson06} with those based on 
the inferred neutral hydrogen, to ensure our comparisons with 
empirical data have not been biased in any obvious manner.  As the 
current implementation of ISM physics within \textsc{Gasoline} does not 
allow gas to cool to densities more appropriate to molecular hydrogen, 
what we label as `HI' can possess column densities $\sim$5-10$\times$ 
larger than encountered in nature.  For that reason, one must be careful 
to not overinterpret the quoted surface densities.

\subsection{Radial Density Profiles} 
\label{subsec:1} 

\begin{figure}[JENAMscalelengths] 
\includegraphics[scale=.5]{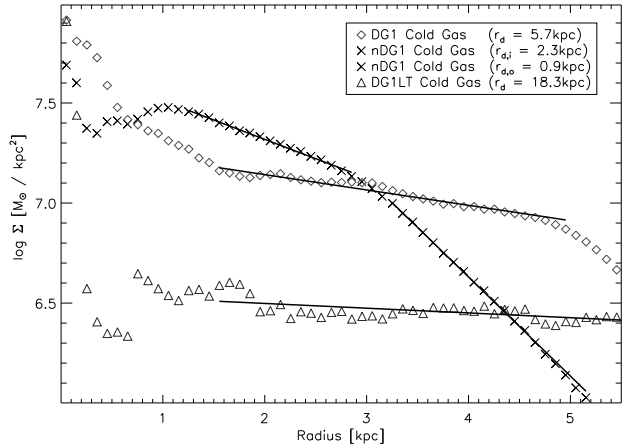} 
\caption{Radial cold gas density profiles for the simulated dwarfs DG1 
(diamonds), DG1LT (triangles), and nDG1 (crosses). The thick overplotted 
lines show the exponential fits to the distributions, from which the 
noted scalelengths were derived. nDG1 is best represented by a double 
exponential, with a break between the two near $\sim$3~kpc. The cold gas 
of DG1 is distributed in a more extended exponential disk component of 
scalelength $\sim$6~kpc, while that of DG1LT is $\sim$18~kpc.}
\label{scalelengths} 
\end{figure}

\vspace{-5mm}
We first examine the radial density profiles 
of the cold gas in the three simulations (Fig~\ref{scalelengths}). 
Observations of HI disks \cite{Tam09,Obr10} show that, in nature, they extend 
from $\sim$2 to $\sim$6 radial disk scalelengths and truncate near 
$\sim$1.5r$_{d}$. DG1LT shows a flat profile with a formal scalelength of 
$\sim$18~kpc, which truncates at $\sim$0.5r$_{d}$. The
base simulation (DG1) has a radial 
scalelength of $\sim$6~kpc (which is also approximately
where it truncates); embedded within
this extended disk is a high density cold gas core. 
Our new simulation (nDG1) 
can really only be fit using two exponentials, spanning the inner
and outer disks (such double exponentials are fairly common in nature),
and also possesses a compact cold core.  These cores are 
somewhat transient, being disrupted during periods of more significant 
star formation; e.g., for DG1, at $z$$\sim$0.7 (the last epoch of 
substantial star formation - 0.1M$_{\odot}$/yr compared to 0.005M$_{\odot}$/yr
at the present-day), the cold core is absent.  These conclusions
are robust to the choice of `HI', as opposed to `cold gas'.

\subsection{Flaring} 
\label{subsec:2} 

\begin{figure}[JENAMScaleHeight] 
\sidecaption 
\includegraphics[scale=.55]{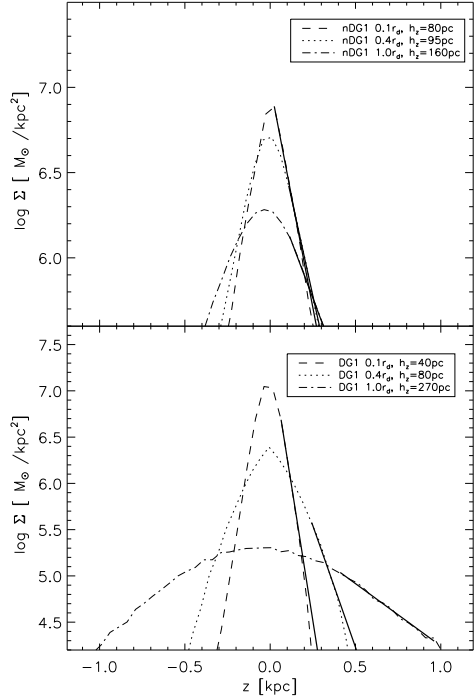} 
\caption{Vertical density profiles of the cold gas disks for nDG1 (upper 
panel) and DG1 (lower panel), where `cold' here means gas particles with 
temperatures less than 15000~K. The profiles are measured in three 
seperate annuli corresponding to 0.1, 0.4, and 1.0 HI disk scale 
lengths. Overlaid on the plots are the exponential fits from which we 
derived the associated scaleheights (inset to each panel).}
\label{scaleheight} 
\end{figure}

\vspace{-5mm}
Both DG1 and nDG1, when viewed edge-on, show significant flaring of 
their gas disks. In nature, comparable dwarfs \cite{Obr10} typically show 
an increase in the FWHM of their vertical density distribution of 
$\sim$50$\%$ in going radially from 0.4r$_{d}$ to 1.0r$_{d}$. 
Fig~\ref{scaleheight} shows the vertical density profiles for the 
simulations at three annuli (0.1r$_{d}$, 0.4r$_{d}$, and 1.0r$_{d}$); we 
can see that DG1 flares by a factor of $\sim$4 from 0.4r$_{d}$ to 
1.0r$_{d}$, while nDG1 flares by a factor of $\sim$1.5. In that 
`fractional' sense, DG1 seems somewhat extreme, but 
quantifying the flaring in terms of physical units (kpc), the degree of 
flaring is not dissimilar to that observed \cite{Obr10}.

\subsection{Velocity Dispersion} 
\label{subsec:3} 

\begin{figure}[JENAMVeldisp]
\includegraphics[scale=.5]{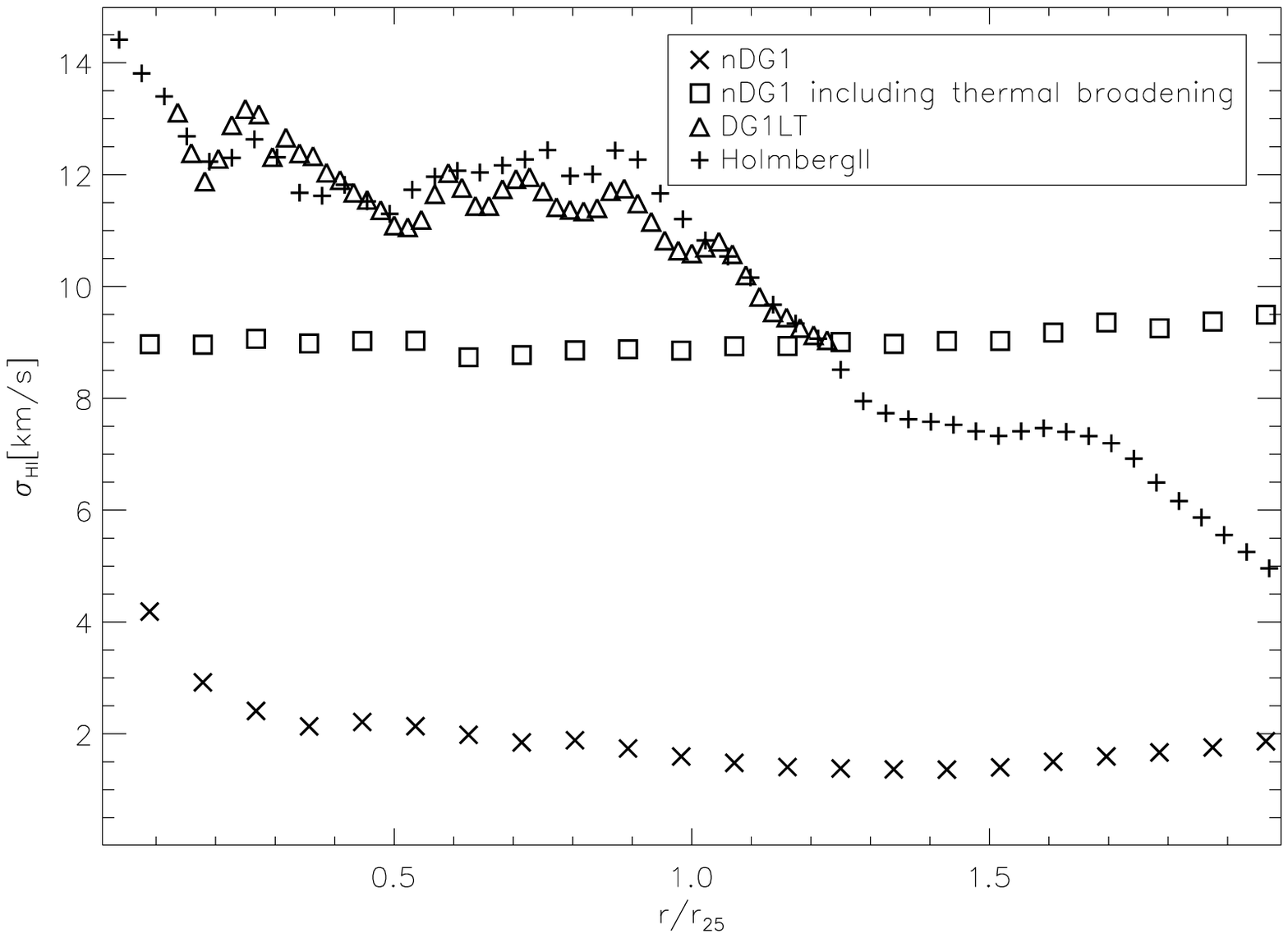}
\caption{HI line-of-sight (40$^\circ$ viewing angle) 
velocity dispersion profiles 
of nDG1 (crosses: particle streaming motions only), 
nDG1 with the inclusion
of thermal broadening (squares), DG1LT (triangles), and 
observational data for Holmberg~II (plus signs), scaled to $r_{25}$ in the
B-band (isophotal radius corresponding to 25~mag/arcsec$^2$, roughly the 
extent of the star forming disk).}
\label{VelDisp}     
\end{figure}

\vspace{-5mm}
Observations of HI 
velocity dispersion profiles \cite{Tam09}, for both giants and dwarfs,
show that galaxies possess characteristic dispersions of
$\sim$8$-$10km/s (rising to $\sim$12$-$15~km/s in the inner star 
forming part of the disk). In Fig~\ref{VelDisp}, we show the 
HI line-of-sight velocity dispersion 
profiles for nDG1, DG1LT, and Holmberg~II \cite{Tam09}.
Critically, we show nDG1 with (squares) \it and \rm without (crosses) the 
inclusion of thermal broadening, following the methodology of
\cite{VDB02}. Because SPH tracks only the streaming motions of the gas
particles, one must incorporate the random velocity
component to each particle, using the internal energy of each particle
as input.  In practice, we draw random velocities for each
Cartesian coordinate from a Gaussian with 
$\sigma = \sqrt{kT/\mu}$ and add those to the components of the streaming
motion.  Without the inclusion of thermal 
broadening, nDG1 (and DG1) shows an extremely (and unphysically) kinematically
cold ISM compared to that of DG1LT (which
despite its aforementioned problems, resembles
empirical data such as HoII very well).  Two unresolved issues which are
apparent from Fig~\ref{VelDisp} are that (i) while the 
dispersion based upon the streaming velocities (crosses) shows the 
same increase in the star forming portion of the disk that is 
observed in nature \cite{Tam09}, the magnitude of the 
thermal broadening essentially ``wipes out'' this signal, and
(ii) the velocity ellipsoid of the cold gas becomes isotropic, 
disguising any anisotropies that might have been present in the
streaming motions (i.e., young stars, and the cold gas from which they
formed, will necessarily have different velocity ellipsoids).

When DG1 is viewed at redshift $z$$\sim$0.7, the more turbulent ISM
being driven by the enhanced and sustained star formation
(0.1M$_{\odot}$/yr compared to 0.005$_{\odot}$/yr), was a close match 
to the velocity dispersion profiles observed in nature - put another
way, the dispersion of the streaming motions during periods of 
significant star formation are already on the order of the dispersions 
observed in nature, without the application of thermal broadening.
As the gas clouds out of which stars are forming within these simulations
have temperatures on the order of $\sim$8000K, there is no way to avoid
the imposition of a significant $\sim$8~km/s thermal broadening term; 
until we can resolve densities corresponding to cores of molecular clouds, 
this SPH limitation must be observed.

\subsection{Power Spectrum}
\label{subsec:4}

\begin{figure}[JENAMpowerspec]
\includegraphics[scale=.5]{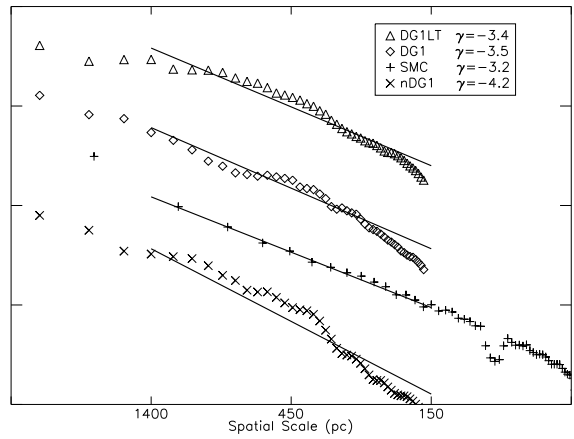}
\caption{Power spectra of the HI distributions of the three simulations -
nDG1 (crosses), DG1 (diamonds), and DG1LT (triangles) - accompanied by 
that for the SMC.  We truncated the power spectra for the simulations
at $\sim$200pc (twice the beam size used to simulate the effects of 
beam smearing).  The break in the SMC spectrum is 
due to a missing base line \cite{Stan99}.  Overplotted on each dataset
is a representative power law fit to the data, with the slopes
noted in the inset to the panel.}
\label{Powerspectrum}       
\end{figure}

\vspace{-5mm}
We next derived the spatial power spectra of the HI gas 
associated with our simulations, as well as that for the Small
Magellanic Cloud \cite{Stan99}, as shown in
Fig~\ref{Powerspectrum}, using
Fourier Transforms of the moment zero density maps, after \cite{Stan99}.
In the simulated galaxies, we smoothed with a 100pc Gaussian to be
consistent with the 
beam smearing present in the observational data. 
All of the galaxies were then fit with a 
power law $P$$\propto$$k^{\gamma}$, where
$\gamma$$=$$-$3.5 for DG1, $\gamma$$=$$-$3.4 for DG1LT,
$\gamma$$=$$-$4.2 for nDG1, and $\gamma$$=$$-$3.2 for the SMC. 
The steeper slope possessed by nDG1 indicates an excess of power 
distributed on large scales within its simulated ISM, relative to that
observed in the other simulations or the SMC (i.e., the enhanced 
feedback has shifted power
from smaller scales to larger ones).  The observational data
(for the SMC and other dwarfs) is well fit by a pure power law, 
indicative of a lack of preference in HI cloud size, in nature),
while the simulations show departures from this scenario. These
departures can be traced to specific structures in the simulations - 
e.g., in nDG1, the enhanced power seen on scales of $\sim$400$-$500pc
corresponds to the spacing of the 
tightly-wound spiral structures seen in its inner
few kpcs.

\section{Chemistry of the Simulated Dwarfs}
\label{sec:3}

\vspace{-5mm}
The next direction for our work will be based upon a comparison of the
stellar chemistry of the simulations with existing observational data.
To date, we have examined the metallicity
distribution functions (MDFs) and oxygen-to-iron distributions.  Here, we
simply highlight preliminary results pertaining to nDG1.

\begin{figure}[JENAMcumulative]
\includegraphics[scale=.5]{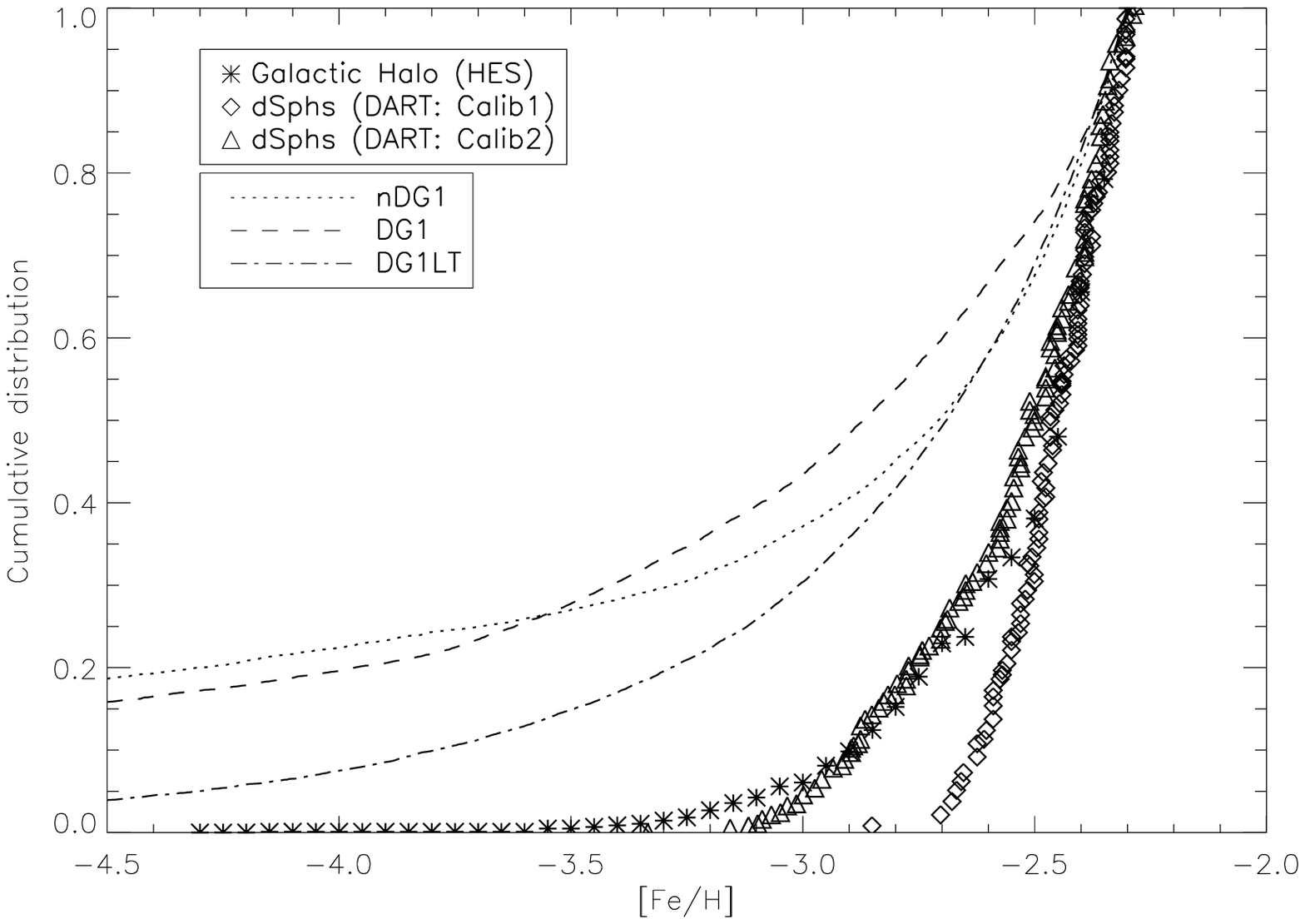}
\caption{Cumulative MDFs for nDG1 (dotted line), DG1 (dashed line) and DG1LT
(dot dashed line). The compilation of observational data shown \cite{Scho09}
includes that of the Galactic halo (asterisks) and the combined MDF
constructed by merging those of the Carina, Sextans, and Sculptor dwarfs
from DART data using two different metallicity calibrations (triangles and
diamonds).}
\label{cumulative}       
\end{figure}

Fig~\ref{cumulative} shows the present-day cumulative
[Fe/H] MDF for all stars within 4~kpc of the three simulations, normalised
to unity (arbitrarily) at [Fe/H]=$-$2.3.
Focusing on the metal-poor tail, it is apparent that
our simulated dwarfs suffer from
a ``G-dwarf problem'' (relative overproduction
of metal-poor stars), an issue to which we will return in an
upcoming study.

The global MDFs of the simulations are similar to those seen in 
Local Group dwarfs, with an $\sim$0.16~dex dispersion `peak' at
[Fe/H]$\approx$$-$1.1 superimposed on a broader
metal-poor component (the formal dispersion of entire nDG1 
MDF being $\sim$0.32~dex). This 
is not dissimilar to the $\sim$0.25~dex (intrinsic)
dispersion seen (for example) 
in the Local Group dwarf, Carina \cite{koch06}.  While
nDG1 is not a dwarf spheroidal, it does have an episodic star formation
history akin to that of Carina. 

\begin{figure}[JENAMOFendg1_sculp]
\includegraphics[scale=.5]{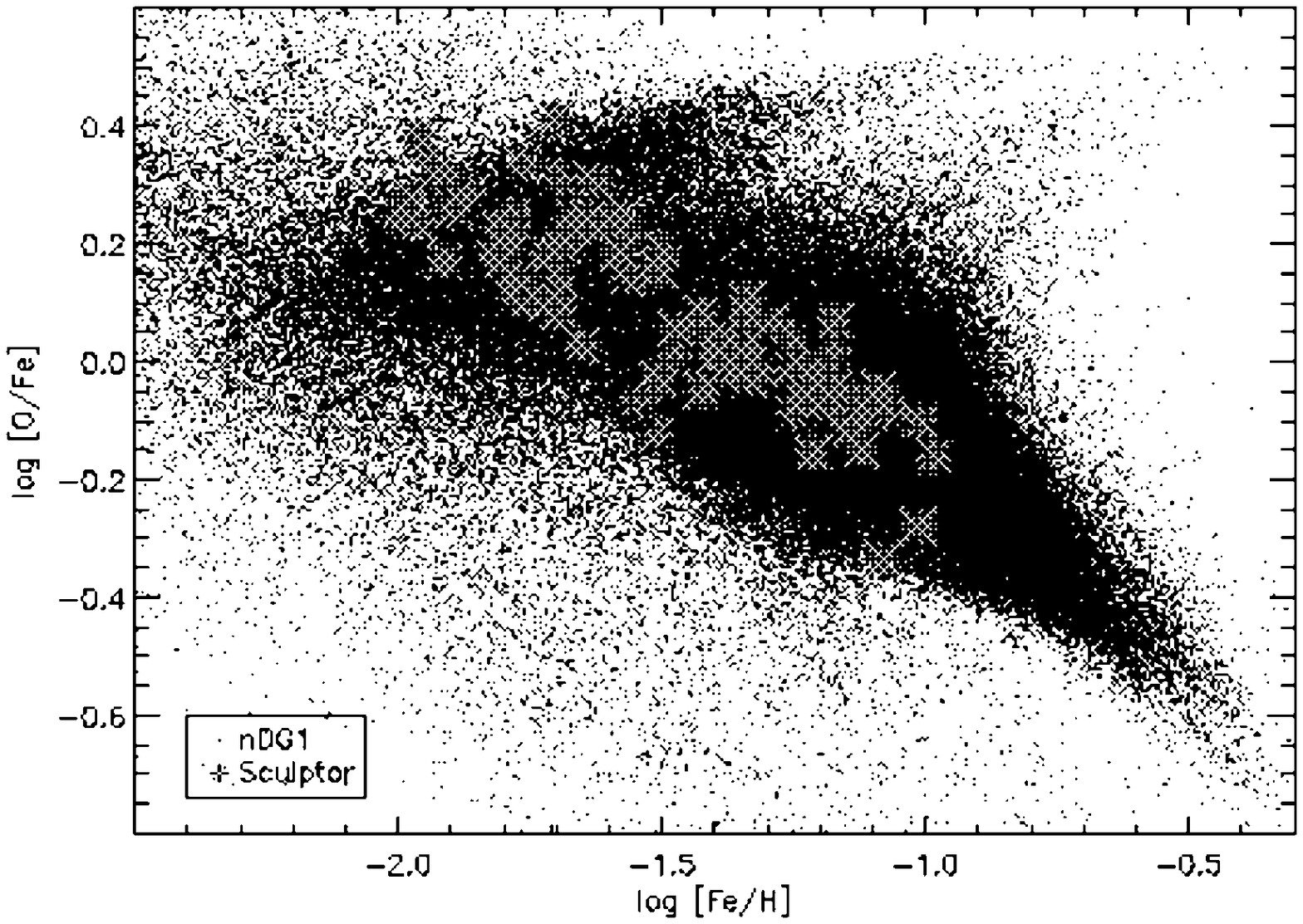}
\caption{[O/Fe]--[Fe/H] distribution for all stars within 4kpc of nDG1.
For comparison, data from the Local Group dwarf, Sculptor\cite{Tol09}, 
are overplotted (asterisks).}
\label{OFe}       
\end{figure}

Fig~\ref{OFe} shows the relationship between [O/Fe] and [Fe/H] for the
same stars (black dots) within 4~kpc of nDG1. Overplotted, again not 
because it is supposed to be a true analog of nDG1 but because it
provides a useful benchmark, are the 
[$\alpha$/Fe]-[Fe/H] data (asterisks)
for the Local Group dwarf, Sculptor \cite{Tol09}.
The distribution of the residuals in [O/Fe] about the best-fit lines
through the Sculptor and nDG1 datasets are both consistent with intrinsic
scatters of $\sim$0.13~dex. For nDG1, this scatter varies somewhat
with metallicity, with the scatter in the [O/Fe] residuals for stars near 
[Fe/H]$\approx$$-$1 being $\sim$0.1~dex, while those with [Fe/H]$<$$-$1.5
show a scatter closer to $\sim$0.2~dex. To first order though, it would 
suggest that the adopted 
magnitude of metal diffusion employed was reasonable

%

\end{document}